\documentclass[useAMS,usenatbib]{mn2e}

\usepackage{graphicx}
\usepackage{color}
\usepackage{latexsym}
\usepackage{lscape}
\usepackage{longtable}

\title[SiO emission from a dense cold core]
{Detection of SiO emission from a massive dense cold core}

\author[N. Lo et al.]{N.~Lo,$^{1,2}$\thanks{E-mail: nlo@phys.unsw.edu.au}
M.~Cunningham,$^{1}$ I.~Bains,$^{1,3}$ M. G.~Burton$^{1}$
and G.~Garay$^{4}$\\
$^{1}$School of Physics, University of New South Wales, Sydney, NSW 2052,
Australia\\
$^{2}$Australia Telescope National Facility, CSIRO, PO Box 76, Epping, NSW
1710, Australia\\
$^{3}$Centre for Astrophysics and Supercomputing, Swinburne University of Technology,
P.O. Box 218, Hawthorn, VIC 3122, Australia\\
$^{4}$Departamento de Astronom\'ia, Universidad de Chile, Casilla 36-D, Santiago, Chile\\}

\begin{document}

\date{Accepted ***. Received ***; in original form ***}

\pagerange{\pageref{firstpage}--\pageref{lastpage}} \pubyear{}

\maketitle

\label{firstpage}

\begin{abstract}
We report the detection of the SiO (J = 2 $\to$ 1) transition from the massive
cold dense core G333.125$-$0.562. The core remains undetected at wavelengths
shorter than 70-$\mu$m and has compact 1.2-mm dust continuum. The SiO emission
is localised to the core. The observations are part of a continuing
multi-molecular line survey of the giant molecular cloud G333. Other detected
molecules in the core include $^{13}$CO, C$^{18}$O, CS, HCO$^+$, HCN, HNC,
CH$_3$OH, N$_2$H$^+$, SO, HC$_3$N, NH$_3$, and some of their isotopes. In
addition, from NH$_3$ (1,1) and (2,2) inversion lines, we obtain a temperature
of 13 K. From fitting to the spectral energy distribution we obtain a colour
temperature of 18 K and a gas mass of 2 $\times$ 10$^3$ M$_{\scriptsize \sun}$.
We have also detected a 22-GHz water maser in the core, together with methanol
maser emission, suggesting the core will host massive star formation. We
hypothesise that the SiO emission arises from shocks associated with an outflow
in the cold core.
\end{abstract}


\section{Introduction}

\subsection{Dense cold core G333.125$-$0.562}
The massive, dense, cold core G333.125$-$0.562 ($\alpha_{J2000} = 16^h21^m35^s,
\delta_{J2000} = -50^d41^m10^s$) is located in the G333 giant molecular cloud
complex, at a distance of 3.6-kpc. It has a dust mass of 2.3 $\times$ 10$^3$
M$_{\scriptsize \sun}$ and an average density of 2 $\times$ 10$^5$ cm$^{-3}$
\citep{2004ApJ...610..313G}. The cold core was discovered by
\citet{2004ApJ...610..313G}, comparing the SEST Imaging Bolometre Array (SIMBA)
1.2-mm survey \citep{2004A&A...426...97F} with \textit{Midcourse Space
Experiment (MSX)} mid-infrared data and \textit{Infrared Astronomical Satellite
(IRAS)} far-infrared data. \citeauthor{2004ApJ...610..313G} noted strong 1.2-mm
continuum emission from this source in the absence of emission in any of the
\textit{MSX} and \textit{IRAS} bands. From the 1.2-mm dust continuum, along
with the upper limits of \textit{IRAS} fluxes, they inferred that the core is
extremely cold ($<$ 17 K), massive and dense. \citeauthor{2004ApJ...610..313G}
hypothesised that the core is at an early stage of star formation, most likely
before the development of an internal
luminosity source, and will collapse to form high-mass star(s).\\
\indent \citet{2002A&A...384L..15P} conducted SIMBA 1.2-mm observations of the
source based on the detection of a class II 6.6-GHz methanol maser from survey
by the Mt Pleasant Observatory \citep{1996MNRAS.280..378E}.
\citeauthor{2002A&A...384L..15P} suggested that there is a very deeply embedded
object because of the detection of strong 1.2-mm emission and the methanol
maser. Later, a class I 95.1-GHz methanol was detected by
\citet{2005MNRAS.359.1498E} with the Mopra Telescope.

\section{Observations and Data Processing}
The multi-molecular line data presented here were collected between July 2004
and October 2006 with the Mopra Telescope, operated by the Australia Telescope
National Facility (ATNF). 
It has a full width half-maximum (FWHM) beam size of $\sim$32 arcsec at 100-GHz
\citep{2005PASA...22...62L}. The observations were carried out with the new
UNSW Mopra Spectrometer (MOPS) digital filterbank back-end and Monolithic
Microwave Integrated Circuit (MMIC) receiver, except for $^{13}$CO and
C$^{18}$O which were observed with the Mopra Correlator (MPCOR) and the
previous Superconductor Insulator Superconductor
(SIS) receiver \citep{2006MNRAS.367.1609B}. 
According to \citet{2005PASA...22...62L} the main beam efficiency at 86-GHz is
0.49, and at 115-GHz is 0.42.\\
\indent The observations are part of the on-going multi-molecular line survey
(the `Delta Quadrant Survey' or DQS) project carried out by the University of
New South Wales, of the giant molecular cloud complex G333. The cold core
G333.125$-$0.562 is part of the complex. The first
paper featuring $^{13}$CO is already published \citep{2006MNRAS.367.1609B}. 
In addition to the 3-mm multi-molecular line survey, NH$_3$ and H$_2$O (12-mm)
maps of the core were also obtained during December 2006 with the Mopra
Telescope. The main beam efficiency for 12-mm system is $\eta _\mathrm{22 GHz}
= 0.7$ and FWHM beam size approximately 2 arcmin. The data presented in this
work has a pointing accuracy of within $\sim$ 5 arcsec.

\section{Results and discussions}
\begin{figure}
  \centering
  \includegraphics[]{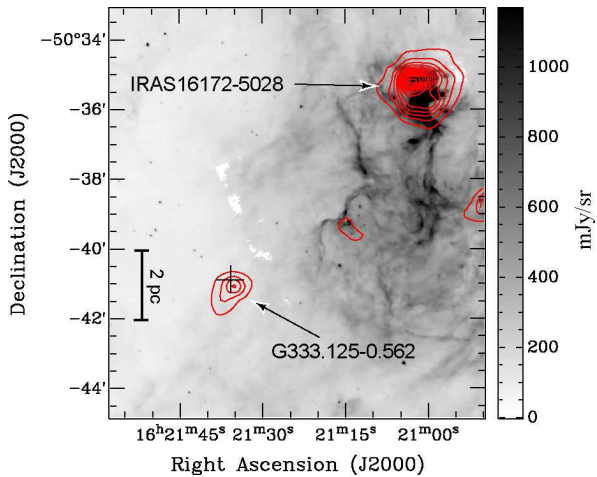}
  \caption[]{GLIMPSE 8-$\mu$m image (grey scale) overlaid with the SIMBA 1.2-mm
  dust continuum (red contours). The contour levels start from 690 mJy
  beam$^{-1}$ with increments of 690 mJy beam$^{-1}$. The arrows show the position
  of the core and the IRAS source, the cross shows the 6.7-GHz methanol maser,
  while the scale bar denotes 2-pc at a distance of 3.6-kpc.
  No 8-$\mu$m emission is evident at the position of the core.}
\label{fig:SIMBA_8um}
\end{figure}

\begin{figure}
  \centering
  \includegraphics[]{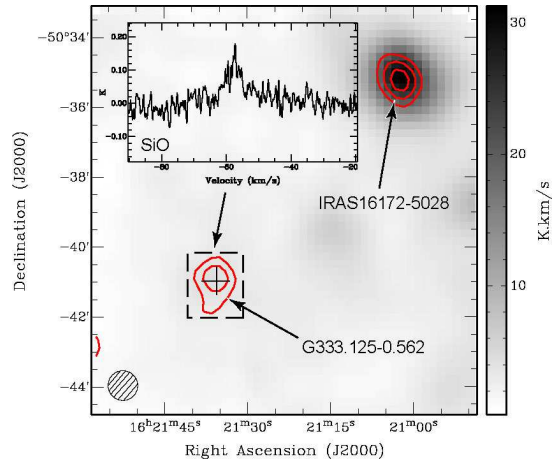}
  \caption[]{The integrated emission map of SiO $J = 2 \to 1$ transition
  (red contours) overlaid on CS $J = 2 \to 1$ transition (grey scale). The
  contour levels start from 2.4 K km$^{-1}$ in steps of 0.8 K km$^{-1}$ (1
  $\sigma$), in terms of the main beam brightness temperature,
  $T\mathrm{_{MB}}$. The cross shows the 6.7-GHz methanol maser, the hatched
  circle indicates the beam size. The inset is the velocity profile of SiO
  averaged over the region indicated by the dashed box. Note the temperature
  scale of the SiO spectra is in antenna temperature.}
\label{fig:G333-SiO}
\end{figure}

Shown in Figure \ref{fig:SIMBA_8um} is the \textit{Spitzer} Galactic Legacy
Infrared Midplane Survey Extraordinaire (GLIMPSE) 8-$\mu$m
image 
(grey scale) overlaid with the SIMBA 1.2-mm dust
continuum (red contours) obtained from \citet{2004A&A...426..119M}. From the
image it can be seen that the core (bottom left) is isolated and has compact
dust emission. No infrared emission is evident from it. Inspection of the
\textit{Spitzer} MIPSGAL images
at 24- and 70-$\mu$m show a source only at the longer
wavelength. Thus this must be a cold dust core. From this core we have further
detected emission from a number of molecular transitions, their observed
parameters are summarised in Table \ref{tab:lines_phys}. 
Among these, the detection of SiO (J $=2 \to 1$) is of particular interest.
Presented in Figure \ref{fig:G333-SiO} is the integrated emission map of SiO
(red contours) overlaid on CS (grey scale), both maps have been smoothed to 36
arcsec. From the map we can see there are two strong peaks of SiO emission, one
is the luminous IRAS source IRAS16172$-$5028 and the other one is from the
cold, dense core G333.125$-$0.562. This is quite different to the extended
distribution seen in other molecular transitions in this region. Shown in the
inset is the SiO velocity profile averaged over the core (dashed box), it is
evident the line profile is broad. We will first discuss the distribution of
the molecules, followed by the physical properties derived for the core.

\begin{figure*}
  \includegraphics[]{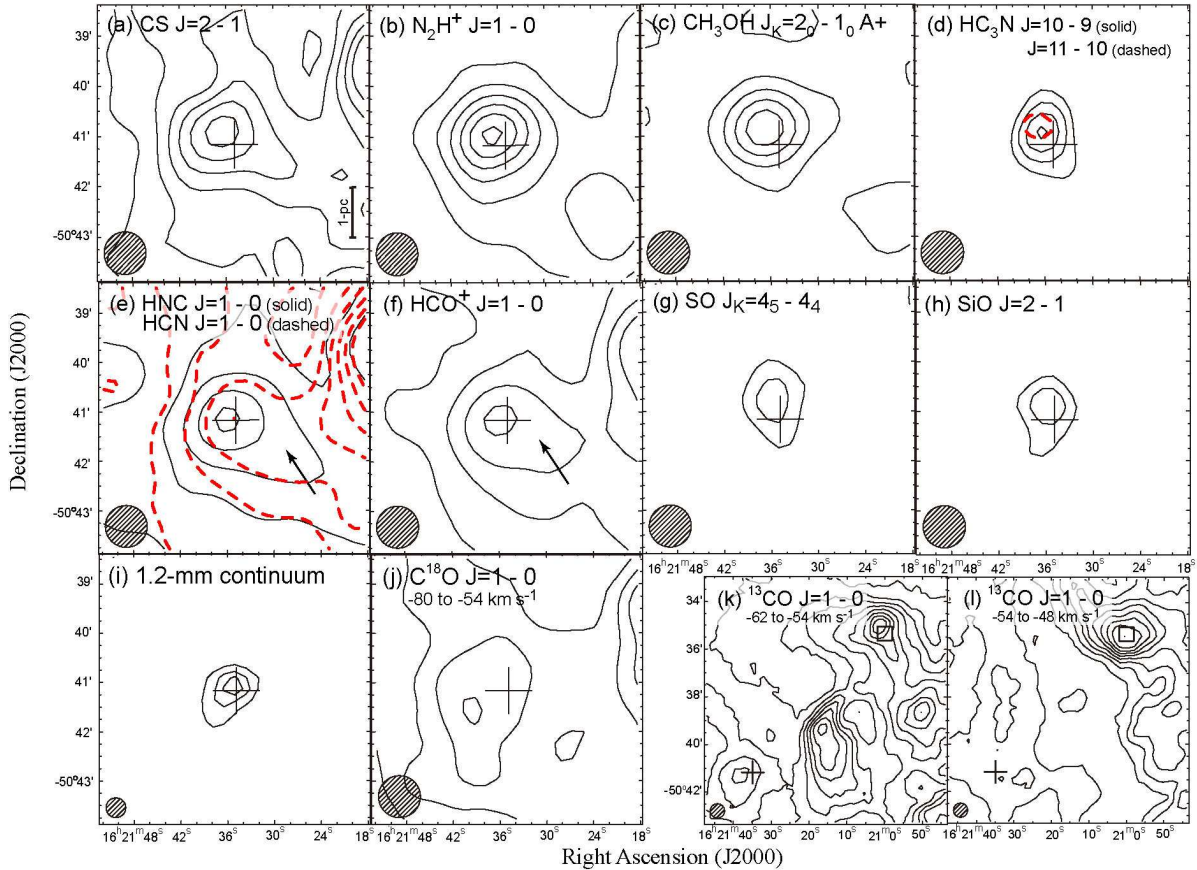}
  \caption[]{ (a) to (j): Integrated emission maps of various molecular lines,
  as indicated for the core G333.125$-$0.562.
  The contour levels start from 3 $\sigma$ of the r.m.s. The contour levels of
  N$_2$H$^+$, CH$_3$OH, HNC, HCN, HCO$^+$, C$^{18}$O and 1.2-mm
  continuum are in steps of 3 $\sigma$, and 1 $\sigma$ for CS, HC$_3$N
  (both transitions), SO and SiO. The r.m.s. for CS and CH$_3$OH is 1.1 K km
  s$^{-1}$, N$_2$H$^+$ is 0.91, HNC, HCO$^+$ and SiO are 0.82 K km
  s$^{-1}$, HC$_3$N (J $= 10 -9$) and HCN are 0.61 K km s$^{-1}$,
  HC$_3$N (J $=11 -10$) and SO is 0.91 K km s$^{-1}$, and C$^{18}$O is
  1.2 K km s$^{-1}$. The temperatures are in $T\mathrm{^*_{MB}}$. The integrated
  velocity range is from $-80$ to
  $-40$ km s$^{-1}$, except C$^{18}$O which is from $-80$ to $-54$
  km s$^{-1}$. The arrows show the tongue extended emission in the south-west
  direction.
  (k): $^{13}$CO integrated emission map from $-62$ to $-54$ km s$^{-1}$, over a
  region covers the core (lower left) and IRAS16172-5028 (upper right). (l): Same as
  figure k but over velocity range $-54$ to $-48$ km s$^{-1}$. Contour levels for
  $^{13}$CO maps start from 20 per cent of the peak, in increments of 10 per
  cent. The cross marks the peak of the 1.2-mm continuum and the square is the
  UCH\texttt{II} region associated with IRAS 16172$-$5028. The beam sizes are
  indicated with the hatched circle, the scale bar denotes 1-pc at distance of 3.6-kpc.}
\label{fig:int_emission}
\end{figure*}

\subsection{Distribution of the molecules}
Presented in Figure \ref{fig:int_emission} are the integrated emission (zeroth
moment) maps of the detected molecular lines over velocity range of $-80$ to
$-40$ km s$^{-1}$, except $^{13}$CO and C$^{18}$O, which are integrated over
the ranges indicated in the figure. The molecular line emission peaks are
coincident (within a beam) with the 1.2-mm dust, except for C$^{18}$O which is
offset by approximately 55 arcsec ($\sim$ 1-pc at a distance of 3.6-kpc). Note
that there is extended C$^{18}$O emission towards the north of the dust peak.
The low density tracer CO is clearly more extended than the other molecules,
which are sensitive to denser gas. Recent studies \citep{2005ApJ...620..795L}
suggest that CS is a good tracer of dense cores in high-mass star forming
regions in contrast to low-mass star formation where CS may be depleted. Since
the peak matches well with the dust peak, this suggests that CS does not suffer
from significant depletion. However, we do note that depletion could happen at
smaller scales. Another feature evident in the HNC and HCO$^+$ maps is the
tongue of extended emission to the south-west (arrow). A comparison between HCN
and HNC integrated emission maps shows large differences in distribution, HCN
being more extended than HNC, while at the same time the peak brightness of HCN
is lower than HNC. Whether this is caused by optical depth effects needs
further investigation. Although HCN and HNC have the same precursors (HCNH$^+$
and H$_2$CN$^+$) and similar dipole moments, it is believed that HCN is
enhanced in warm environments in contrast to HNC, which forms preferentially in
colder conditions \citep{2006ApJS..162..161K}. \citet{1998ApJ...503..717H}
found the abundance ratio of [HNC]/[HCN] rapidly drops as the temperature
exceeds a temperature of 24-K. N$_2$H$^+$ is known as a cold gas tracer as its
abundance tends to be enhanced in cold dense cores, due to the depletion of CO,
which decreases the destruction of N$_2$H$^+$ and H$_3^+$
\citep*{{1992ApJ...387..417W},{2004A&A...415.1065H}}. Rather than reacting with
CO to form HCO$^+$ and H$_2$, the H$_3^+$ ion reacts with N$_2$ to form
N$_2$H$^+$. From the integrated emission map of N$_2$H$^+$, we see a compact
distribution of N$_2$H$^+$. This suggests the N$_2$H$^+$ is tracing the outer
cold envelope of the dense core. The other two molecules which show a compact
distribution are CH$_3$OH and HC$_3$N (both transitions), both of which trace
dense gas. SiO and SO are known to be greatly enhanced in outflows and shocked
regions \citep*{1992A&A...254..315M}. Detecting both these species suggests
that energetic activity associated with shock waves is present.

\begin{table*}
  \centering
  \caption[]{Summary of the observed and derived parameters of the molecular lines:
  centre velocity (\emph{V}), line width (\emph{$\Delta$V}), peak brightness
  ($T\mathrm{_{peak}}$), integrated brightness ($\int{T\mathrm{_{MB}}\mbox{ }dv}$),
  isotopic column density (\emph{N}), virial mass ($M\mathrm{_{vir}}$) and FWHM
  angular size ($\theta$). The second last column is the abundance ratio
  (\emph{X}) of the molecular line relative to H$_2$. For lines with more than one
  velocity component, parameters reported here are for the component associated with
  the core velocity, $-58$ km s$^{-1}$. The uncertainty range is in parentheses.}
  \begin{minipage}[]{1\textwidth}
  \begin{tabular}{c c c c c c c c c c c} 
    \hline
    \hline
    Molecule & Transition & \emph{V} & $\Delta$\emph{V}
        & $T\mathrm{_{peak}}$ & $\int{T\mathrm{_{MB}}\mbox{ }dv}$
        & \emph{N} & $M\mathrm{_{vir}}$ & $\theta$ & \emph{X} & Note\\
     & & \scriptsize (km s$^{-1}$) & \scriptsize (km s$^{-1}$) & \scriptsize (K)
        & \scriptsize (K km s$^{-1}$) & \scriptsize ($10^{14}$ cm$^{-2}$)
        & \scriptsize ($10^{3}$ M$_{\sun}$) & \scriptsize (arcsec)
        & \scriptsize ($10^{-10}$) & \\
    \hline
    \textbf{3-mm}\\
    $^{13}$CO & $1 - 0$ & $-$57.56 (0.03) & 4.90 (0.06) & 9.5 & 49.3 (0.8)
        & 2.0 $\times 10^3$ & 1.2 & 48 & $5.0 \times 10^4$ & [ex] \\
    C$^{18}$O & $1 - 0$ & $-$56.63 (0.05) & 5.1 (0.1) & 2.6 & 14.1 (0.3)
        & 5.2 $\times 10^3$ & 1.3 & 48 & $1.3 \times 10^4$ & [ex] \\
    CS & $2 - 1$ & $-$58.8 (0.3) & 4.5 (0.7) & 0.9 & 4.1 (1.3)
        & 1.7 & 1.0 & 48 & 4.3 & [ex] \\
    C$^{34}$S & $2 - 1$ & $-$57.7 (0.3) & 3.1 (0.8) & 0.3 & 1.0 (0.2)
        & 0.6 & 0.5 & 54 & 1.5 & [c] \\
    HCO$^+$ & $1 - 0$ & $-$59.4 (0.1) & 4.6 (0.1) & 2.1 & 10.4 (0.4)
        & 2.7 & 1.1 & 48 & 6.8 & [ex] \\
    HCN & $1 - 0$ & & & & & & & & & [hf] \\
     & F $=2-1$ & -59.5 (0.1) & 2.2 (0.3) & 0.6 & 1.4 (0.2) & $-$ & $-$
        & $-$ & $-$ & $-$ \\
     & F $=0-1$ & -66.6 (0.1) & 5.3 (0.4) & 0.7 & 3.9 (0.3) & $-$ & $-$
        & $-$ & $-$ & $-$ \\
     & F $=1-1$ & -51.8 (0.4) & $-$ & $-$ & $-$ & $-$ & $-$
        & $-$ & $-$ & [I]\\
    HNC & $1 - 0$ & -59.07 (0.05) & 3.9 (0.1) & 2.2 & 9.2 (0.5)
        & $>3 \times 10^{-3}$ & 0.8 & 46 & $>9 \times 10^{-3}$ & [II] \\
    N$_2$H$^+$ & $1 - 0$ & $-$58.8 (0.1) & 4.0 (0.1) & 1.0 & 3.9 (0.3)
        & 24 & 3.3  & 99 & 61 & [hf,c]\\
    CH$_3$OH & $2_{02} - 1_{01}$ A+ & $-$58.0 (0.1) & 4.2 (0.2)
        & 1.2 & 5.2 (0.3) & $-$ & $-$ & 84 & $-$ & [c,III] \\
    HCCCN & $11 - 10$ & $-$57.3 (0.2) & 3.6 (0.5) & 0.3 & 1.3 (0.2)
        & 0.6 & 1.3 & 69 & 1.5 & [c,IV]\\
    HCCCN & $10 - 9$ & $-$57.41 (0.09) & 3.8 (0.2) & 0.6 & 2.5 (0.1)
        & $-$ & $-$ & $-$ & $-$ & [c]\\
    SO & $3_2 - 2_1$ & $-$57.2 (0.2) & 5.0 (0.5) & 0.5 & 2.7 (0.2) & 1.1
        & 1.3 & 48 & 2.6 & [II] \\
    SiO & $2 - 1$ v=0 & $-$57.2 (0.3) & 9.0 (1.0) & 0.2 & 1.8 (0.2)
        & 4.4 & 3.9 & 48 & 11 & [II]\\
    \hline
    \textbf{12-mm}\\
    NH$_3$ & $(1,1)$ & $-$56.89 (0.02) & 3.46 (0.03) & 0.9 & 10.7 (0.2)
        & 0.4 & 1.5 & 120 & 1 & [hf,V]\\
    NH$_3$ & $(2,2)$ & $-$57.2 (0.1) & 3.7 (0.2) & 0.4 & 2.7 (0.2)
        & $-$ & $-$ & $-$ & $-$ & [hf]\\
    H$_2$O & $6_{16}-5_{23}$ & $-$53.9 (0.1) & 2.0 (0.3) & 0.1 & 0.24 (0.03)
        & $-$ & $-$ & $-$ & $-$ & [m]\\
    \hline
  \end{tabular}\\
  \begin{tabular}{l p{17cm}}
   \multicolumn{2}{l}{N\tiny OTE \small $-$} \\
   & [hf] stands for hyperfine structures; the peak brightness is for the
     main component, and the integrated brightness for the sum of all
     hyperfine components, except HCN, which we list each hyperfine component.\\
   & [m] stands for maser.\\
   & [ex] indicates the molecular line emission is extended, hence we used the
     angular size of dust continuum to derive the virial mass.\\
   & [c] for compact emission.\\
   & [el] indicates the emission is elongated, the angular size is the average
     of RA and DEC.\\
   & [I] hyperfine component is blended with the $-50$ km s$^{-1}$ velocity
     component.\\
   & [II] we derived the column density assuming the line is optically thin, hence it
     and the abundance ratio are lower limits.\\
   & [III] no other CH$_3$OH transition detected, therefore column density cannot
     be calculated.\\
   & [IV] column density calculated using the rotational temperature (5.7-K)
     obtained from $J = 10 \to 9$ and $J = 11 \to 10$ lines.\\
   & [V] the beam size at 22-GHz is $\sim2$ arcmin, therefore the source is
     unresolved. We took the angular size to be one beam width as there is
     no noticeable extended emission.\\
  \end{tabular}
  \end{minipage}
  \label{tab:lines_phys}
\end{table*}

\subsection{Physical properties}
In order to derive the physical parameters, spectra were spatially averaged
over the FWHM angular size for molecules with compact emission, otherwise over
a 48 arcsec region, then fitting with a Gaussian in \small CLASS\normalsize
\footnote{Continuum and Line
Analysis Single-dish Software, part of GILDAS software package by IRAM.\\
(http://www.iram.fr/IRAMFR/GILDAS/)}. The fitted parameters, i.e. line width,
centre velocity, brightness temperature and optical depth (from hyperfine
structure fitting of N$_2$H$^+$ and NH$_3$) were then used to derive column
densities and masses. The derived parameters are summarised in Table
\ref{tab:lines_phys}. The column densities were calculated with an excitation
temperature of 15-K, which is the average of the dust and NH$_3$ values (see
below). We note that the derived column densities vary by 20 per cent for the
excitation temperature range ($13 - 18$ K). We have derived molecular abundances
relative to the H$_2$ column density, which is obtained from the 1.2-mm dust
continuum, $N\mathrm{_{H_2} \sim 4 \times 10^{23}}$ cm$^{-2}$.\\
\indent From the NH$_3$ (J,K) = (1,1) and (2,2) inversion transitions we have
calculated the rotational temperature ($T\mathrm{_{12}}$) of the core,
following the method stated in \citet*{1986A&A...157..207U}. We assumed NH$_3$
(1,1) and (2,2) are tracing the same volume of gas, and all hyperfine
components have the same excitation temperature. With the above assumptions, we
have derived the rotational temperature, $T_{12} = 13.1$ K, with an uncertainty
range of $12.9 - 13.3$ K. According to \citet{1988MNRAS.235..229D}, below 20-K
the rotational temperature of NH$_3$ (1,1)-(2,2) matches the kinetic
temperature ($T\mathrm{_{kin}}$) well.\\
\indent We are also able to determine a dust temperature, mass and luminosity
from the spectral energy distribution (SED). Inspection of the GLIMPSE and
MIPSGAL images of the core only shows evidence of emission at the longest
wavelength (70-$\mu$m). Far-IR balloon-borne measurements have been made at 150
and 210-$\mu$m by \citet{2001MNRAS.326..293K}, through a 3 arcmin beam.
Examination of the \emph{Spitzer} infrared images show this emission must be
confined to the G333.125$-$0.562 core. Hence we use the 150-, 210-$\mu$m and
1.2-mm fluxes to determine the SED, constraining the size by the angular size
(48 arcsec) of the 1.2-mm core, and applying a greybody fit with a dust
emissivity index of $\beta = 2$ \citep{2006MNRAS.368.1223H}. This yields the
following parameters for the core: $T\mathrm{_{dust}}=18.6 \pm 0.1$ K,
$L_{\mathrm bol} = (8.5 \pm 0.3) \times 10^3$ $\mathrm{L_{\scriptsize \sun}}$
and $M\mathrm{_{gas}=1.8 \times 10^3}$ $\mathrm{M_{\scriptsize \sun}}$. Taken
with the angular size these yield \={n}$\mathrm{_{H_2} \sim 1.3 \times 10^5}$
cm$^{-3}$ and $N\mathrm{_{H_2} \sim 4 \times 10^{23}}$ cm$^{-2}$. We note these
parameters are similar to those derived by \citet{2004ApJ...610..313G} who made
use of just the 1.2-mm flux, combined with upper limits at 60 and 100-$\mu$m
from IRAS data $-$ yielding an upper limit to the temperature of 17-K. Thus,
the dust-determined temperature is comparable to that derived from the NH$_3$
emission, giving us confidence in the conclusion that the core is exceedingly
cold. If we assume that the molecular lines are thermalized then
$T\mathrm{_{kin}}$ is approximately equal to the excitation temperature
($T\mathrm{_{ex}}$) for each molecule.\\
\indent Most of the detected molecular lines are peaked at $\sim$ $-$57 km
s$^{-1}$, except for HCO$^+$, HNC and N$_2$H$^+$ which are peaked at $\sim$
$-$59 km s$^{-1}$. This likely due to lines being optically thick, for instance
the H$^{13}$CO$^+$ line peaks at $-$57 km s$^{-1}$. The methanol maser also
peaks at $-$57 km s $^{-1}$, with components at $-$53 and $-$63 km s $^{-1}$
\citep*{2005MNRAS.359.1498E}, indicative of outflows. The water maser is at
$-$54 km s$^{-1}$. The line widths for most of the molecules are between 3 to 5
km s$^{-1}$ as opposed to the $\sim 0.5$ km s$^{-1}$ predicted for quiescent
gas dominated by thermal broadening, suggesting the core is turbulent. The SiO
however has the largest line width amongst the detected molecular lines, 9.0
$\pm$ 1.1 km s$^{-1}$, leading further support to being associated with an
outflow. Its abundance with respect to H$_2$ has an upper limit of $\sim$
10$^{-12}$ in quiescent gas, but is greatly enhanced in powerful shocks, up to
$\sim$ 10$^{-7}$ \citep*[e.g.][]{1989ApJ...343..201Z,1992A&A...254..315M}. We
have determined the abundance to be $\sim 10^{-9}$, in between these extremes.
A similar abundance ratio has been found in molecular outflows, e.g.
\citet*{1999A&A...343..585C}, \citet{2002ApJ...567..980G}. A follow up CO (J $=
1 \to 0$) molecular line observation over a five square arcmin region of the
core was taken in June 2007; the spectra clearly shows line wings, however they
are confused with multiple velocity components and need further investigation.

\section{What is generating the SiO emission: outflows or
collisions?} 
\begin{figure}
  \centering
  \includegraphics[]{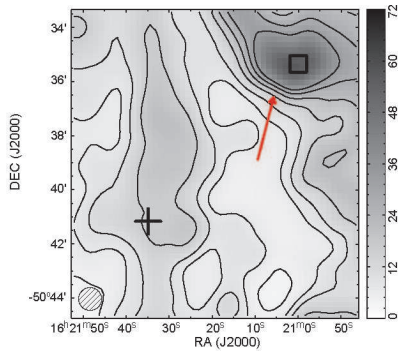}
  \caption[]{The integrated emission map of $^{13}$CO (J = 1 $\to$ 0) from
  $-52$ to $-50$ km s$^{-1}$. The contour levels are 1.4
  , 4.3, 7.1,
  10, 21, 27 and 33 K km s$^{-1}$. The temperature is in terms of the
  main beam brightness temperature. The cross marks the cold core
  G333.125$-$0.562 and the square marks IRAS 16172$-$5028. The arrow indicates
  a likely region where the gas is being compressed.}
\label{fig:13CO_non_cc}
\end{figure}

The optically thick isotopomers clearly show two different
velocity components, indicating there are two separate clouds in this region.
The cloud at $\sim -50$ km s$^{-1}$ is associated with IRAS 16172$-$5028, and
is at the ambient velocity of G333 giant molecular cloud (see Figure
\ref{fig:int_emission}l). The other cloud at $\sim -58$ km s$^{-1}$ is
associated with the cold core G333.125$-$0.562 (Figure
\ref{fig:int_emission}k). The SiO emission could be generated by these two
clouds colliding, but we consider this unlikely. First, consider the confined
emission of the SiO, localised to the dust core itself. If it is due to a
cloud-cloud collision we would expect more extended distribution. Secondly, for
a cloud-cloud collision there should be distinctive features in the density
structure. Shown in Figure \ref{fig:13CO_non_cc} is the $^{13}$CO integrated
emission map of velocity range $-52$ to $-50$ km s$^{-1}$. Note the sharp edge
(arrow) on the left side of the IRAS source (square), suggesting compression of
the gas here. However no compression is evident associated with the cold core
(cross) at the velocity observed.

\section{Summary and Conclusion}
From our molecular line survey of the G333 Giant Molecular Cloud, we have
detected thermal SiO emission 
from the massive, cold, dense core, G333.125$-$0.562, with an abundance
enhanced over typical unshocked molecular cloud values. This core has a gas
mass of 1.8 $\times$ 10$^3$ M$_{\scriptsize \sun}$, is undetected up to
70-$\mu$m and has compact 1.2-mm dust continuum emission. From the NH$_3$ 
inversion lines we derived a temperature of 13 K, which is comparable with that
derived from the SED (19 K). The detection of compact emission from a cold gas
tracer (N$_2$H$^+$) suggests this is a dense cold core. Typical line widths are
between 4 to 5 km s$^{-1}$, indicating the core is turbulent, as expected in
massive star formation.\\
\indent In conclusion, from these observations we believe that the cold massive
core harbours a deeply embedded, massive protostellar object that is driving an
outflow. This is occurring at a very early stage of star formation, prior to
the creation of an infrared source in the core.


\section*{Acknowledgments}
The Mopra Telescope is part of the Australia Telescope and is funded by the
Commonwealth of Australia for operation as National Facility managed by CSIRO.
The University of New South Wales Digital Filter Bank used for the observations
with the Mopra Telescope was provided with support from the Australian Research
Council. We would like to thank the reviewer's constructive comments on
improving this paper, T. Wong for his work on C$^{18}$O data, and S. N.
Longmore for help in analysing the NH$_3$ data. This research (GLIMPSE and
MIPSGAL images) has made use of the
NASA/IPAC Infrared Science Archive which is operated by the Jet Propulsion
Laboratory, California Institute of Technology, under contract with NASA. The
molecular line rest frequencies are from \citet*{1979ApJS...41..451L}.

\bibliographystyle{mn2e}
\bibliography{letter1}


\bsp

\label{lastpage}

\end{document}